\documentclass[aps,prl,twocolumn,nofootinbib,groupedaddress,amsfonts,floatfix]{revtex4}
\usepackage{graphicx,amsmath,amssymb,amstext,amssymb,amsbsy,amsfonts,amsthm,lscape,units}

\usepackage[usenames,dvipsnames]{color}

\newcommand{\gws}{GWs}
\newcommand{\gdw}{GW}
\newcommand{\Gdw}{GW}
\newcommand{\Gws}{GWs}

\usepackage{color}
\usepackage{ifthen}
\newboolean{editorial}
\setboolean{editorial}{true}
\newcommand{\editorial}[2]{\ifthenelse{\boolean{editorial}}{\textcolor{red}{[\textsf{\textbf{{#1}}}: }\textcolor{blue}{\textsf{{#2}}}\textcolor{red}{]}}{}}

\begin{document}

\title{The detectability of cosmological gravitational-wave backgrounds: a rule of thumb}

\author{John T. Giblin, Jr${}^{1,2}$}
\email{giblinj@kenyon.edu}
\author{Eric Thrane${}^3$}
\email{ethrane@ligo.caltech.edu}

\affiliation{${}^1$Department of Physics, Kenyon College, Gambier, OH 43022 USA}
\affiliation{${}^2$Department of Physics, Case Western Reserve University, Cleveland, OH 44106, USA}
\affiliation{${}^3$LIGO Laboratory, California Institute of Technology, MS 100-36, Pasadena, CA, 91125 USA}

\begin{abstract}
  The recent claim by BICEP2 of evidence for primordial gravitational waves from inflation has focused interest on the potential for early-Universe cosmology using observations of gravitational waves.
    In addition to cosmic microwave background detectors, efforts are underway to carry out gravitational-wave astronomy over a wide range of frequencies including pulsar timing arrays (nHz), space-based detectors (mHz), and terrestrial detectors ($\sim$$10$--$\unit[2000]{Hz}$).
    This multiband effort will probe a wide range of times in the early Universe (each corresponding to a different energy scale), during which gravitational-wave backgrounds may have been produced through processes such as phase transitions or preheating.
    In this letter, we derive a rule of thumb (not quite so strong as an upper limit) governing the maximum energy density of cosmological backgrounds.
    For most cosmological scenarios, we expect the energy density spectrum to peak at values of $\Omega_\text{gw}(f)\lesssim10^{-12\pm2}$.
    We discuss the applicability of this rule of thumb and the implications for gravitational-wave astronomy.
\end{abstract}

\maketitle

Pending confirmation, the detection by BICEP2 of primordial gravitational waves (GWs) has launched the era of \gdw\ astronomy~\cite{bicep2}.
Cosmic microwave background detectors observe \gws\ at low frequencies $f\lesssim\unit[10^{-16}]{Hz}$.
The direct detection of \gws\ by other observatories, and at much higher frequencies, is likely imminent.
In the coming years, a network of terrestrial \gdw\ detectors~\cite{aligo,virgo,kagra,geo}, operating at $\sim$$10$--$\unit[2000]{Hz}$, is expected to detect dozens of compact binary coalescences per year~\cite{cbc_rates}.
Pulsar timing arrays~\cite{Hobbs-et-al:2010}, operating at $\sim$$\unit[5]{nHz}$, are approaching the sensitivity required for the \gdw\ detection from super-massive black hole binaries~\cite{pta_science}.
Space-based detectors~\cite{bbo,decigo}, operating at $\unit[0.1]{mHz}$--$\unit[1]{Hz}$, are all but guaranteed to observe \gws\ from e.g., merging supermassive black holes.

Astrophysical backgrounds can arise, e.g., from super-massive black hole binaries~\cite{jaffe}, stellar-mass binaries~\cite{StochCBC}, white dwarf binaries~\cite{phinney_whitedwarfs}, neutron stars~\cite{RegPac}, the first stars~\cite{firststars}, and---more speculatively---cosmic (super)strings \cite{Berezinsky:2000vn}.
Cosmological backgrounds, created by major events in the history of the Universe, can be created from or following inflation~\cite{grishchuk,starob,peloso}, by phase transitions~\cite{Kosowsky:1992rz,Kamionkowski:1993fg}, and from alternative cosmologies~\cite{PBBpaper}.
While astrophysical backgrounds are interesting in their own right, cosmological backgrounds are especially interesting since they shed light on fundamental physics.
Here, we focus on cosmological backgrounds.

Recent observations by BICEP2~\cite{bicep2} of a primordial background are most simply explained as the amplification of vacuum fluctuations following inflation~\cite{guth}, implying a background with a nearly flat energy density spectrum:
\begin{equation}\label{eq:Omega}
  \Omega_\text{gw}(f) \equiv \frac{1}{\rho_c}\frac{d\rho_\text{gw}}{d\ln f} .
\end{equation}
Here, $\rho_\text{gw}$ is \gdw energy density, $f$ is frequency, and $\rho_c$ is the critical energy density required for a closed Universe.
Extrapolating the BICEP2 measurement to the LIGO band, we expect $\Omega_\text{gw}(f)\approx10^{-15}$.
\Gdw\ backgrounds at this level are too weak to observe directly except by the most ambitious detectors~\cite{bbo} and, of course, using the cosmic microwave background~\cite{bicep2}.

However, cosmological signals, produced through other mechanisms, can produce considerably more detectable signals with $\Omega_\text{gw}\lesssim10^{-12}$.
Detection of a cosmological background above the level predicted for the amplification of vacuum fluctuations could point to a richer and more interesting early Universe than posited by the simplest version of slow-roll inflation.

Here, we draw attention to generic features common to many cosmological backgrounds in order to derive a ``rule of thumb" governing the maximum likely amplitude of most cosmological backgrounds.
We use the phrase ``rule of thumb'' rather than ``upper limit'' to convey the theoretical uncertainty in our derivation.
The rule of thumb employs assumptions consistent with a large number of models in order to provide a broadly (if not universally) applicable prediction governing the maximum energy density of cosmological backgrounds.  The point is to provide a systematic framework for understanding trends among predictions of cosmological backgrounds.

Our rule of thumb applies to backgrounds (at all energy scales) created after inflation during the radiation-dominated epoch.   
During this epoch, the age of the Universe varied between $\unit[10^{-35}]{s}\lesssim t\lesssim\unit[47000]{yr}$, corresponding to energy scales of $\unit[10^{15}]{GeV}\gtrsim E \gtrsim\unit[1]{eV}$, and redshifts $z\gtrsim3500$ \cite{Ade:2013zuv}.
We assume that the length scale of the source is smaller than the cosmological horizon $H^{-1}$, which follows from causality.
Our rule of thumb does {\em not} apply to astrophysical backgrounds, which can peak well above cosmological models~\cite{StochCBC}, as they are created at much later times.

We assume that the background evolves as a Friedmann-Lem\"aitre-Robertson-Walker spacetime.
\Gws\ propagate as strain perturbations $h_{ij}$ in synchronous gauge,
\begin{equation}
\label{sync}
ds^2 = dt^2 - a^2(t)\left[\delta_{ij} + h_{ij}\right]dx^idx^j,
\end{equation}
where the propagating degrees of freedom are the two polarizations that obey the transverse-traceless conditions,
\begin{equation}
h_i^i = 0\,\,\,{\rm and} \,\,\,\,h_{i,j}^j =0.
\end{equation}
These strain perturbations obey sourced Klein-Gordon equations,
\begin{equation}
  \ddot{h}_{ij} + 3H \dot{h}_{ij} - \frac{1}{a^2} \nabla^2 h_{ij} =( 16 \pi  G)S^{TT}_{ij} ,
\label{equationofmotion}
\end{equation}
where the source is the transverse-traceless {\em projection} of the anisotropic stress tensor,
\begin{equation}\label{aniso}
  S_{ij}^{TT} = T_{ij} - \frac{\delta_{ij}}{3}T_k^k.
\end{equation}

Our objective is to estimate $\Omega_\text{gw}(f)$ from a relatively generic cosmological source.
To this end, we link \gdw\ energy density $\rho_\text{gw}$ to the energy density of some source $\rho_\text{gw} < \rho_s $, which, in turn, represents some fraction of the total energy density in the Universe $\rho_s < \rho$.
By considering the fraction of energy density available for the source, and the fraction of the source energy density converted to \gws, we estimate the maximum $\Omega_\text{gw}(f)$ today.

We make a few assumptions about the source.
We consider a source associated with a characteristic scale $k_*$, and assume that components of the stress-energy tensor can be written in momentum space as
\begin{equation}\label{eq:Gaussian}
  \tilde{T}_{ij} (\vec{k}) \approx \tilde{T} (\vec{k})=  
  A
  \exp \left[-\frac{(|\vec{k}|-k_*)^2}{2\sigma^2}\right],
\end{equation}
where each $T_{ij}(k)$ is approximately the same magnitude, $\sigma$ parameterizes the source width, and $A$ is the peak height.
Although $A$ is determined by the detailed physics of each source, it cannot exceed the total energy density of the Universe at the time of the process.

The isotropic pressure of the source, 
\begin{equation}
\tilde{p}_{s}  (\vec{k})=\frac{1}{3}\left(\tilde{T}_{11} (\vec{k})+\tilde{T}_{22} (\vec{k})+\tilde{T}_{33} (\vec{k})\right) = \tilde{T} (\vec{k}),
\end{equation}
is related to the energy density of the source,
\begin{equation}
\tilde{\rho}_s (\vec{k}) = \frac{ \tilde{p}_s (\vec{k}) }{w} = \frac{ \tilde{T} (\vec{k})}{w}
\end{equation}
by $w$, which relates the magnitude of the stress-energy tensor of the source to the source energy density.
If we chose a volume large enough so the configuration-space energy density is homogeneous, we can use Parseval's theorem to relate the momentum space energy spectrum to the total source energy in a volume $V$,
\begin{equation}
  \int d^3k \left|\tilde{\rho}_s(\vec{k})\right|^2 =  \int dV \rho^2_s (\vec{x})\approx  V \rho^2_s .
\end{equation}

We define
\begin{equation}
  W(k_*,\sigma) \equiv 
4\pi 
  \int_0^\infty k^2 \exp\left[-\frac{(k-k_*)^2}{\sigma^2}\right] \,dk .
\end{equation}
The magnitude of the stress-energy tensor and the source energy density are related:
\begin{equation}\label{eq:Asq}
  \left|A\right|^2 = \frac{w^2\rho_s^2 V}{W(k_*,\sigma)} .
\end{equation}
The \gdw\ energy created in this process is only a fraction, $\alpha <1$ of the total energy budget of the Universe: $\rho_s = \alpha {\rho}$.
Thus,
\begin{equation}
  \left|\tilde{T}\right|^2 = \frac{w^2\alpha^2V\rho^2}{W(k_*,\sigma)}\exp\left[-\frac{(k-k_*)^2}{\sigma^2}\right] .
\end{equation}

Next, we calculate the size of the metric perturbations.  Since each mode $h_{ij}(\vec{k})$ obeys a sourced Klein-Gordon equation (assuming that the source is short-lived compared to the Hubble time, allowing us to momentarily ignore the Hubble Friction term), we estimate the maximum size of $\tilde{h}_{ij}$ by studying the point when the acceleration of $h_{ij}(\vec{k})$ vanishes.
In the language of a harmonic oscillator, we evaluate the size of the metric perturbation by balancing the force due to the source with the restoring force.
It follows that the $h_{ij}$ are approximately the same:
\begin{equation}
  \label{heq}
  \tilde{h} \approx \tilde{h}_{ij} \approx \frac{16 \pi G}{k^2}S^{TT} .
\end{equation}

Last, we relate the size of the transverse-traceless anisotropic stress tensor to the size of the stress-energy tensor:
\begin{equation}
  \label{sourcerel}
  \beta \equiv \frac{\left|S^{TT}\right|^2}{\left|T\right|^2} .
\end{equation}
The projection of $T_{ij}$ onto $S_{ij}^{TT}$ extracts the tensor-part of the stress-energy tensor and is therefore sensitive to the source geometry.
This is the hardest parameter to estimate without specific knowledge of the source.

We determine the magnitude of $A(\vec{k})$, but not the phase, necessary to estimating $\beta$.
One realization of $A_{ij} = \left|A\right|e^{i\theta_{ij}}$ has six independent phases.
We randomly chose six phases and project the stress-energy tensor $T_{ij}$ onto the transverse-traceless anisotropic stress tensor $S^{TT}_{ij}$ generating a distribution of $\beta$.
Using a simulation, we determine $\overline\beta\approx 10^{-1.5}-10^{-2}$ for a random process.

We use 
\begin{equation}
  \Omega_\text{gw}(k) =
  \frac{1}{\rho}\frac{k^3}{32\pi G} \frac{1}{V} \sum_{i,j} \int d\Omega\,
  \Bigl| \dot{h}_{ij}^{\rm TT}(t,\mathbf{k}) \Bigr|^2
  \label{omega}
\end{equation} 
to calculate $\Omega_\text{gw}(k)$ at the time when the source vanishes~\cite{Easther:2006vd}.
We exchange numerical factors for the sum in Eq~\ref{omega} and evaluate the angular part of the integral,
\begin{equation}\label{eq:exchange_num_facs}
\sum_{ij} \int d\Omega \left|\dot{\tilde{h}}_{ij}\right|^2 = 36 \pi  \left|\dot{\tilde{h}}\right|^2
\end{equation}
where
\begin{equation}
  \label{dothrel}
  \left|\dot{\tilde{h}}\right|^2  =\left|\tilde{h}\right|^2 k^2  = (16 \pi G)^2\frac{ \beta\left|\tilde{T}\right|^2}{k^2}
\end{equation}
via Eqs.~\ref{heq} and \ref{sourcerel}.
We combine Eq.~\ref{dothrel} with Eq.~\ref{eq:exchange_num_facs} and plug into Eq.~\ref{omega}, yielding
\begin{equation}
  \Omega_{\rm gw}(k) =\frac{288\pi^2 G}{\rho V}k\beta\left|\tilde{T}\right|^2  = \frac{108 \pi}{\rho^2 V}H^2 k \beta^2 \left|\tilde{T}\right|^2 ,
\end{equation} 
where the final equality follows from Friedmann's equation:
\begin{equation}
  H^2 = \frac{8\pi G}{3} \rho .
\end{equation}
When the source vanishes,
\begin{equation}\label{eq:Omg108}
  \Omega_{\rm gw}(k) = 108\pi \alpha^2 \beta w^2 \frac{H^2k}{W(k_*,\sigma)} e^{\left[-(k-k_*)^2/\sigma^2\right]} .
\end{equation}

It might seem surprising that we can write this spectrum so simply.
In particular, Eq~\ref{eq:Omg108} depends on the dimensionless quantity $H^2k/W(k_*,\sigma)$, and so we need not know the scale $k_*$.
The peak energy density can be estimated by evaluating
\begin{equation}
  \Omega_{\rm gw} (k_*) \approx 108\pi \, \alpha^2 \beta w^2 N(k_*,\sigma) .
\end{equation}

Last,
\begin{equation}\label{eq:N}
  N(k_*,\sigma) \equiv \frac{H^2k_*}{W(k_*,\sigma)} = \frac{(k_*H^{-1})}{W(k_*H^{-1},\sigma H^{-1})} .
\end{equation}
We investigate Eq.~\ref{eq:N} numerically.
For fixed $k_*$, $N(k_*,\sigma)$ diverges as $\sigma \rightarrow 0$, but only for unphysically small values of $\sigma$.
Generally, the source width can be a few orders of magnitude smaller than the characteristic frequency.
Nonetheless, decreasing the source width changes the amplitude of the \gdw\ spectrum modestly.
For small values of $\sigma/k_*$,
\begin{equation}
  N(k_*,\sigma)\propto \left(\frac{k_*}{\sigma}\right) .
\end{equation}

In practice, we expect that $\sigma < k_*$, so we estimate $N(k_*,\sigma)$ by setting the ratio of $\sigma/k_*$.
In the small $\sigma/k_*$ limit,
\begin{equation}
  N(k_*,\sigma) \rightarrow 0.0449\left(\frac{k_*}{\sigma}\right)\left(\frac{H}{k_*}\right)^2 ,
\label{numericalexpansion}
\end{equation}
where the proportionality constant is obtained evaluating $W(k_*,\sigma)$ numerically.

In present times~\cite{Easther:2006vd,Price:2008hq},
\begin{equation}
  \Omega_{\rm gw,0}(k) h^2 = \Omega_{\rm rad,0}h^2 \big(g_0/g_e\big)^{1/3}\Omega_{\rm gw}(k)
\label{omega1}
\end{equation}
where $0$ indicates the present, $h$ is the dimensionless Hubble parameter, $\Omega_{\rm rad,0}$ is the current energy density from radiation.
The factor $g_0/g_{\rm e}$ is the ratio of the number of degrees of freedom today to the number at matter-radiation equality.
We approximate $g_0/g_{\rm e}=1/10$, recognizing that $(g_0/g_{\rm e})^{1/3}$ changes by only a factor of $\sim$two for every factor of ten in $g_0$.
We adopt $h=0.68$~\cite{Ade:2013zuv} to facilitate comparisons with the observational literature.

Since $\Omega_{\rm rad,0} = 7.78\times10^{-5}$ \cite{Ade:2013zuv} (assuming $N_{\rm eff} - 1 \approx 2$ species of relativistic neutrinos today), we obtain
\begin{equation}
  \Omega_\text{gw,0 } \approx 0.012 \alpha^2 \beta w^2 N(k_*,\sigma) \frac{k}{k_*} e^{\left[-(k-k_*)^2/\sigma^2\right]} .
\label{fullspectrum}
\end{equation}
The frequency today is related to the wave vector at the time of \gdw\ production~\cite{Easther:2006vd,Easther:2006gt,Amin:2014eta}:
\begin{equation}
  f = 6\times10^{10} \frac{k}{\sqrt{m_\text{pl}H}} \,{\rm Hz} =  1.7\times10^{11} \left(\frac{k}{H}\right)\frac{\rho^{1/4}}{m_{\rm pl}} \,{\rm Hz} .
\label{peakfrequency}
\end{equation}
The present-day \gdw\ frequency is a function only of the energy scale at the time of the source $H$ and the dimensionless constant $k/H$.

There are three tunable parameters: $(\alpha,\beta,w)$.
For strong signals $\alpha\lesssim1$, but $\alpha \approx 0.5$ is more likely.
The second parameter, $w\lesssim1$ is likely.
For scalar fields, e.g., $w\approx 1/3$.
The third parameter $\beta$---describing how inherently quadrupolar the energy density is---has the greatest dynamic range.
One can imagine situations in which almost the entire Universe is the source as well as cases in which the source is a small fraction of the energy budget.

Putting everything together, the peak height is
\begin{equation}\label{peak2}
  \begin{split}
    \Omega_\text{gw,0}(k_*)  \approx  2.3\times 10^{-4} \alpha^2 \, \beta \, w^2 \frac{k_*}{\sigma}\left(\frac{H}{k_*}\right)^2 . \\
  \end{split}
\end{equation}
(This scaling is noted in~\cite{1201.0983,0902.2574,0711.2593}.)
Now, we identify plausible values of $k_*/H$ and $\sigma/k_*$.
In principle, $k_*/H$ is different for different cosmological processes.
However, given our goal of constraining the maximum allowable $\Omega_\text{gw}$ from cosmological sources, we chose a value, as small as possible so that $N(k_*,\sigma)$ is as large as possible, subject to constraints from causality: the peak wavelength must be sub-horizon.
Motivated by models of bubble collisions \cite{Kosowsky:1991ua,Kosowsky:1992vn} and phase transitions \cite{Kosowsky:1992rz,Kamionkowski:1993fg,Caprini:2006jb}, we chose fiducial values $k_* = 100 H_*$ and $\sigma/k_*=1/2$.
(While a large class of models employ comparable parameters, other choices can be made for specific models---e.g.,~\cite{1201.0983,0709.2091}---which can be investigated with Eq.~\ref{peak2}.)
We thereby obtain our rule of thumb:
\begin{equation}\label{eq:RuleOfThumb}
  \boxed{
    \Omega_\text{gw,0}(k_*) \approx 4.7\times 10^{-8}\, \alpha^2 \, \beta \, w^2.
  }
\end{equation}
If we repeat the above calculations assuming that $\tilde{T}(\vec{k})$ is described, not by a Gaussian distribution as in Eq.~\ref{eq:Gaussian}, but by a plateau distribution, [constant on $(k_*-\sigma, k_*+\sigma)$ and zero everywhere else], then the resulting rule of thumb prediction is just $9\%$ less.
Thus, the results do not depend strongly on the assumed shape of $\tilde{T}(\vec{k})$.

Equipped with Eq.~\ref{eq:RuleOfThumb}, we consider three different scenarios---corresponding to three sets of tunable parameters $(\alpha,\beta,w)$---reflecting the plausible range of $\Omega_\text{gw}(k_*)$.
These scenarios, described in Tab.~\ref{optimtable}, are labeled ``optimistic,'' ``realistic,'' and ``pessimistic.''
These categorizations, inspired by~\cite{s5_rates}, are necessarily subjective.
However, by providing a range of values, we endeavor to show a range of possible outcomes.
For the realistic scenario, the rule of thumb becomes: $\Omega_{\rm gw,0}(k_*) \approx 1 \times10^{-12}$.

\begin{table}[htb]
  \begin{tabular}{|c|c|c|c|c|}
    \hline 
    {\bf scenario} & $\alpha$ & $\beta$ & $w$ & $\Omega_\text{gw}(k_*)$ \\\hline
    optimistic & $1$ & $0.1$ & $1/3$ & $4.97\times 10^{-10}$ \\ \hline
    realistic & $0.1$ & $0.03$ & $1/3$ & $1.49\times 10^{-12}$ \\ \hline
    pessimistic & $0.02$ & $0.005$ & $1/3$ & $9.93\times10^{-15}$ \\ \hline
  \end{tabular}
  \caption{
    Energy density peak heights for three sets of tunable parameters assuming $\sigma/k_* = 1/2$ and $k_*/H = 100$.
    \label{optimtable}
  }
\end{table}

We now assess the detectability of the three representative rule-of-thumb signals using different \gdw\ detectors.
We consider: (i) Advanced LIGO using $\unit[1]{yr}$ of coincident Hanford-Livingston data at design sensitivity, (ii) the proposed Einstein Telescope using  $\unit[1]{yr}$ of data with the ``ET-D'' sensitivity~\cite{et}, (iii) a hypothetical pulsar timing array from~\cite{locus} consisting of 20 pulsars and assuming $\unit[100]{ns}$ timing noise, $\unit[5]{yr}$ of observation time, and a cadence of $\unit[20]{yr^{-1}}$, and (iv) the Big Bang Observer (BBO)~\cite{Phinney-et-al:2004, Cutler-Harms:2006}, a proposed space-based detector using parameters from~\cite{locus}.

\begin{figure}[htb]
  \includegraphics[width=3.2in]{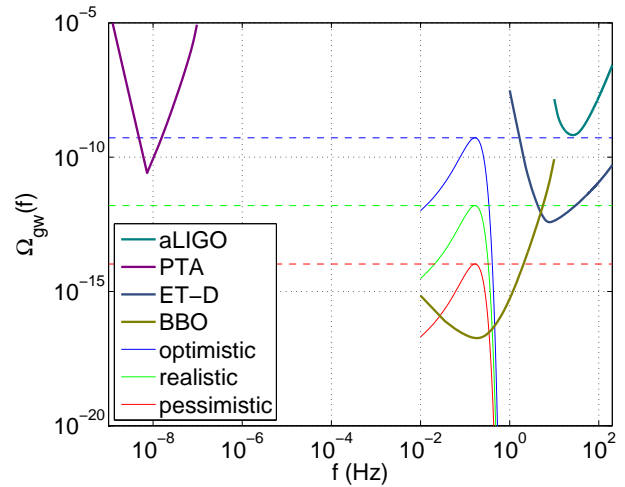}
  \caption{
    Rule-of-thumb energy density spectra compared with sensitivity curves.
    The thick lines are the ``power-law integrated'' sensitivity curves~\cite{locus} (for power-law spectra) for different detectors.
    The thin dashed lines indicate the peak height for our rule of thumb models.
    To avoid clutter, we plot the rule-of-thumb spectra only for one choice of $f_*=\unit[0.15]{Hz}$ (corresponding to the BBO band).
    Any signal spectrum falling entirely below the solid power-law integrated curve will produce a signal-to-noise ratio of $<$1, and is therefore undetectable.
    We ignore the difficulties of astrophysical foregrounds and assume that they have been successfully subtracted.
    We assume a Hubble parameter of $h=0.68$~\cite{Ade:2013zuv}.
  }
  \label{fig:Omgf}
\end{figure}

For each detector, we optimistically tune the peak frequency $f_*\equiv ck_*/2\pi$ to produce the most favorable signal.
The results are summarized in Fig.~\ref{fig:Omgf}.
The rule-of-thumb signals (thin dashed) are compared to the sensitivity curve for each detector (solid).
The sensitivity curves are ``power-law integrated curves''~\cite{locus}, representing the sensitivity of each detector to a broadband stochastic background with a power-law shape.

While the \gdw\ signals we consider here are peaked, not power laws, the power-law integrated curves nonetheless provide a useful guide.
Any dashed rule-of-thumb line falling below the solid power-law integrated curve is undetectable.
Dashed lines intersecting the solid power-law integrated curve might be detectable, and when this happens, we calculate the signal-to-noise ratio (SNR) of a two-detector, cross-correlation search~\cite{Allen-Romano:1999}.

From Fig.~\ref{fig:Omgf}a, all three ($f_*=\unit[23]{Hz}$) rule-of-thumb spectra are out of reach for Advanced LIGO.
The optimistic spectrum can perhaps be probed with additional detectors and/or multiple years of coincident data.
The Einstein Telescope detects a highly significant ($f_*=\unit[6.5]{Hz}$) signal from the optimistic spectrum while the realistic spectrum produces a marginal $\text{SNR}=3.2$ detection.
The pessimistic spectrum is out of reach.
Our hypothetical pulsar timing array unambiguously detects the optimistic ($f_*=\unit[6.8]{nHz}$) spectrum ($\text{SNR}=19$), but not the realistic or pessimistic spectra.
BBO detects statistically significant ($f_*=\unit[0.15]{Hz}$) signatures from all three; $\text{SNR}>380$.
Note: we have ignored complications arising from correlated noise~\cite{schumann_ligo} and the subtraction of astrophysical foregrounds~\cite{paramest,Cutler-Harms:2006}, which may complicate detection.

Our rule of thumb applies to a large subset of cosmological \gdw\ sources that occur {\sl after inflation} and {\sl during the radiation-dominated epoch}.
The argument presented here, after all, relies solely on the ratio of three energy density scales, $\rho_{\rm gw} < \rho_s < \rho$, and on the application of transfer functions.
There are, however, exceptions.

In {\bf non-minimal models of inflation}~\cite{Lopez:2013mqa,Barnaby:2009dd,peloso},  signals evade our bounds because they are ``frozen in" during phase transitions.
\Gws\ remain non-dynamical until they re-enter the horizon when the Universe cools to the appropriate temperature, and so the ratio of energy density scales is irrelevant.
Another possible modification to inflation involves the introduction of direct couplings between the inflaton (usually an axion field) and gauge fields~\cite{Prokopec:2001nc, Anber:2009ua, Barnaby:2011vw}.
As inflation ends, one polarization of the gauge field is dramatically enhanced via a tachyonic process and inflation ends earlier than in canonical slow-roll inflation.
These modes efficiently decay into \gws.

{\bf A non-standard equation of state following inflation} might lead to a detectable cosmological background.
In particular, a ``stiff" equation of state $w>1/3$ modifies the expansion history of the Universe, allowing inflationary gravitational radiation to re-enter the horizon with large amplitudes \cite{Boyle:2007zx}.
This model evades our rule of thumb since the source is not a post-inflationary cosmological process.
However, as pointed out in~\cite{Boyle:2007zx}, there is no theoretical motivation for an effective equation of state larger than $1/3$ after inflation.

If {\bf the graviton is not a massless}, helicity-2 particle (see, e.g. \cite{deRham:2011rn,deRham:2014zqa}) this analysis needs to be rethought in the presence of extra degrees of freedom.
In these models, it is likely that the \gdw\ background would be less diluted when the source is projected, leading, potentially, to a greater value of $\beta\approx1$.
Thus, it may be easier for \gdw\ observatories to detect cosmological backgrounds in non-standard theories of gravity~\cite{Chamberlin:2011ev}.

{\bf Cosmic string networks} produced during phase transitions in the early Universe \cite{Kibble:1976sj} can produce \gws\ in the late Universe via strong bursts of \gws\ produced from cusps~\cite{Berezinsky:2000vn}.
The peak wavelength of these signals is tied to the size of the cosmic strings, not the Hubble scale, and the background is produced at fairly late times (even though the strings themselves are formed very early).
If cosmic string networks were to radiate gravitationally during the radiation-dominated era, these signals would be subject to the constraints presented here, where $k_* \gg H$ (since strings are small compared to the Hubble scale) likely corresponding to very weak signals today.

Our knowledge of the early Universe is far from precise, and \gdw\ astronomy affords us the chance to learn more about this important era.
The coming decades are likely to produce a flood of observational \gdw\ data, which will constrain cosmological models and possibly reveal unknown physics.
As we prepare for this upcoming era of \gdw\ cosmology, it is useful to consider our expectations for what we think we might reasonably detect, based on our present knowledge of the early Universe.
To this end, we have proposed a simple rule of thumb governing the maximum amplitude of cosmological \gdw\ backgrounds: we expect cosmological backgrounds to produce energy density spectra that peak around $\Omega_\text{gw}\approx10^{-12}$.
Our rule of thumb is based on simple scaling arguments and provides robust, if approximate, theoretical guidance for \gdw\ cosmology.

In order to evade the rule-of-thumb assumptions models typically employ assumptions about inflationary dynamics that require some degree of fine tuning.
We argue that, based on our current understanding of the early Universe, the simplest, most natural models predict cosmological \gdw\ backgrounds that follow the rule of thumb.
The rule of thumb does not apply to astrophysical backgrounds since they are created after matter/radiation equality.
Finally, we note that observational constraints from Big Bang nucleosynthesis and the cosmic microwave background limit the integrated energy density of cosmological backgrounds; see, e.g.,~\cite{sendra,allen97}.

JTG is supported by the National Science Foundation, PHY-1068080.
ET is a member of the LIGO Laboratory, supported by funding from United States National Science Foundation.
LIGO was constructed by the California Institute of Technology and Massachusetts Institute of Technology with funding from the National Science Foundation and operates under cooperative agreement PHY-0757058.

\bibliography{giblinthrane_slim_v6}

\begin{thebibliography}{55}
\expandafter\ifx\csname natexlab\endcsname\relax\def\natexlab#1{#1}\fi
\expandafter\ifx\csname bibnamefont\endcsname\relax
  \def\bibnamefont#1{#1}\fi
\expandafter\ifx\csname bibfnamefont\endcsname\relax
  \def\bibfnamefont#1{#1}\fi
\expandafter\ifx\csname citenamefont\endcsname\relax
  \def\citenamefont#1{#1}\fi
\expandafter\ifx\csname url\endcsname\relax
  \def\url#1{\texttt{#1}}\fi
\expandafter\ifx\csname urlprefix\endcsname\relax\def\urlprefix{URL }\fi
\providecommand{\bibinfo}[2]{#2}
\providecommand{\eprint}[2][]{\url{#2}}

\bibitem[{\citenamefont{Ade et~al.}(2014)}]{bicep2}
\bibinfo{author}{\bibfnamefont{P.~A.~R.} \bibnamefont{Ade}}
  \bibnamefont{et~al.} (\bibinfo{year}{2014}),
  \bibinfo{note}{\url{http://arxiv.org/abs/1403.3985}}.

\bibitem[{\citenamefont{{G M Harry (for the LIGO Scientific
  Collaboration)}}(2010)}]{aligo}
\bibinfo{author}{\bibnamefont{{G M Harry (for the LIGO Scientific
  Collaboration)}}}, \bibinfo{journal}{Classical Quantum Gravity}
  \textbf{\bibinfo{volume}{27}}, \bibinfo{pages}{084006}
  (\bibinfo{year}{2010}).

\bibitem[{\citenamefont{{Acernese, F. for the Virgo
  Collaboration}}(2006)}]{virgo}
\bibinfo{author}{\bibnamefont{{Acernese, F. for the Virgo Collaboration}}},
  \bibinfo{journal}{Classical Quantum Gravity} \textbf{\bibinfo{volume}{23}},
  \bibinfo{pages}{S63} (\bibinfo{year}{2006}).

\bibitem[{\citenamefont{{K. Somiya (for the KAGRA
  collaboration)}}(2012)}]{kagra}
\bibinfo{author}{\bibnamefont{{K. Somiya (for the KAGRA collaboration)}}},
  \bibinfo{journal}{Classical Quantum Gravity} \textbf{\bibinfo{volume}{29}},
  \bibinfo{pages}{124007} (\bibinfo{year}{2012}).

\bibitem[{\citenamefont{{H Grote (for the LIGO Scientific
  Collaboration)}}(2008)}]{geo}
\bibinfo{author}{\bibnamefont{{H Grote (for the LIGO Scientific
  Collaboration)}}}, \bibinfo{journal}{Classical Quantum Gravity}
  \textbf{\bibinfo{volume}{25}}, \bibinfo{pages}{114043}
  (\bibinfo{year}{2008}).

\bibitem[{\citenamefont{Abadie et~al.}(2010{\natexlab{a}})}]{cbc_rates}
\bibinfo{author}{\bibfnamefont{J.}~\bibnamefont{Abadie}} \bibnamefont{et~al.},
  \bibinfo{journal}{Classical Quantum Gravity} \textbf{\bibinfo{volume}{27}},
  \bibinfo{pages}{173001} (\bibinfo{year}{2010}{\natexlab{a}}).

\bibitem[{\citenamefont{Hobbs et~al.}(2010)}]{Hobbs-et-al:2010}
\bibinfo{author}{\bibfnamefont{G.}~\bibnamefont{Hobbs}} \bibnamefont{et~al.},
  \bibinfo{journal}{Classical Quantum Gravity} \textbf{\bibinfo{volume}{27}},
  \bibinfo{pages}{084013} (\bibinfo{year}{2010}).

\bibitem[{\citenamefont{Shannon et~al.}(2013)}]{pta_science}
\bibinfo{author}{\bibfnamefont{R.~M.} \bibnamefont{Shannon}}
  \bibnamefont{et~al.}, \bibinfo{journal}{Science}
  \textbf{\bibinfo{volume}{342}}, \bibinfo{pages}{334} (\bibinfo{year}{2013}).

\bibitem[{\citenamefont{Ungarelli et~al.}(2005)\citenamefont{Ungarelli,
  Corasaniti, Mercer, and Vecchio}}]{bbo}
\bibinfo{author}{\bibfnamefont{C.}~\bibnamefont{Ungarelli}},
  \bibinfo{author}{\bibfnamefont{P.}~\bibnamefont{Corasaniti}},
  \bibinfo{author}{\bibfnamefont{R.~A.} \bibnamefont{Mercer}},
  \bibnamefont{and} \bibinfo{author}{\bibfnamefont{A.}~\bibnamefont{Vecchio}},
  \bibinfo{journal}{Classical Quantum Gravity} \textbf{\bibinfo{volume}{22}},
  \bibinfo{pages}{S955} (\bibinfo{year}{2005}).

\bibitem[{\citenamefont{Yagi}(2013)}]{decigo}
\bibinfo{author}{\bibfnamefont{K.}~\bibnamefont{Yagi}}, \bibinfo{journal}{Int.
  J. Mod. Phys. D} \textbf{\bibinfo{volume}{22}}, \bibinfo{pages}{1341013}
  (\bibinfo{year}{2013}).

\bibitem[{\citenamefont{Jaffe and Backer}(2003)}]{jaffe}
\bibinfo{author}{\bibfnamefont{A.~H.} \bibnamefont{Jaffe}} \bibnamefont{and}
  \bibinfo{author}{\bibfnamefont{D.~C.} \bibnamefont{Backer}},
  \bibinfo{journal}{Astrophys. J.} \textbf{\bibinfo{volume}{583}},
  \bibinfo{pages}{616} (\bibinfo{year}{2003}).

\bibitem[{\citenamefont{Wu et~al.}(2012)\citenamefont{Wu, Mandic, and
  Regimbau}}]{StochCBC}
\bibinfo{author}{\bibfnamefont{C.}~\bibnamefont{Wu}},
  \bibinfo{author}{\bibfnamefont{V.}~\bibnamefont{Mandic}}, \bibnamefont{and}
  \bibinfo{author}{\bibfnamefont{T.}~\bibnamefont{Regimbau}},
  \bibinfo{journal}{Phys. Rev. D} \textbf{\bibinfo{volume}{85}},
  \bibinfo{pages}{104024} (\bibinfo{year}{2012}).

\bibitem[{\citenamefont{Farmer and Phinney}(2003)}]{phinney_whitedwarfs}
\bibinfo{author}{\bibfnamefont{A.~J.} \bibnamefont{Farmer}} \bibnamefont{and}
  \bibinfo{author}{\bibfnamefont{E.}~\bibnamefont{Phinney}},
  \bibinfo{journal}{MNRAS} \textbf{\bibinfo{volume}{346}},
  \bibinfo{pages}{1197} (\bibinfo{year}{2003}).

\bibitem[{\citenamefont{Regimbau and de~Freitas~Pacheco}(2001)}]{RegPac}
\bibinfo{author}{\bibfnamefont{T.}~\bibnamefont{Regimbau}} \bibnamefont{and}
  \bibinfo{author}{\bibfnamefont{J.~A.} \bibnamefont{de~Freitas~Pacheco}},
  \bibinfo{journal}{Astron. and Astrophys.} \textbf{\bibinfo{volume}{376}},
  \bibinfo{pages}{381} (\bibinfo{year}{2001}).

\bibitem[{\citenamefont{Sandick et~al.}(2006)\citenamefont{Sandick, Olive,
  Daigne, and Vangioni}}]{firststars}
\bibinfo{author}{\bibfnamefont{P.}~\bibnamefont{Sandick}},
  \bibinfo{author}{\bibfnamefont{K.~A.} \bibnamefont{Olive}},
  \bibinfo{author}{\bibfnamefont{F.}~\bibnamefont{Daigne}}, \bibnamefont{and}
  \bibinfo{author}{\bibfnamefont{E.}~\bibnamefont{Vangioni}},
  \bibinfo{journal}{Phys. Rev. D} \textbf{\bibinfo{volume}{73}},
  \bibinfo{pages}{104024} (\bibinfo{year}{2006}).

\bibitem[{\citenamefont{Berezinsky et~al.}(2000)\citenamefont{Berezinsky,
  Hnatyk, and Vilenkin}}]{Berezinsky:2000vn}
\bibinfo{author}{\bibfnamefont{V.}~\bibnamefont{Berezinsky}},
  \bibinfo{author}{\bibfnamefont{B.}~\bibnamefont{Hnatyk}}, \bibnamefont{and}
  \bibinfo{author}{\bibfnamefont{A.}~\bibnamefont{Vilenkin}}
  (\bibinfo{year}{2000}), \eprint{astro-ph/0001213}.

\bibitem[{\citenamefont{Grishchuk}(1975)}]{grishchuk}
\bibinfo{author}{\bibfnamefont{L.~P.} \bibnamefont{Grishchuk}},
  \bibinfo{journal}{Sov. Phys. JETP} \textbf{\bibinfo{volume}{40}},
  \bibinfo{pages}{409} (\bibinfo{year}{1975}).

\bibitem[{\citenamefont{Starobinskii}(1979)}]{starob}
\bibinfo{author}{\bibfnamefont{A.~A.} \bibnamefont{Starobinskii}},
  \bibinfo{journal}{JETP Lett.} \textbf{\bibinfo{volume}{30}},
  \bibinfo{pages}{682} (\bibinfo{year}{1979}).

\bibitem[{\citenamefont{Barnaby et~al.}(2012)\citenamefont{Barnaby, Pajer, and
  Peloso}}]{peloso}
\bibinfo{author}{\bibfnamefont{N.}~\bibnamefont{Barnaby}},
  \bibinfo{author}{\bibfnamefont{E.}~\bibnamefont{Pajer}}, \bibnamefont{and}
  \bibinfo{author}{\bibfnamefont{M.}~\bibnamefont{Peloso}},
  \bibinfo{journal}{Phys. Rev. D} \textbf{\bibinfo{volume}{85}},
  \bibinfo{pages}{023525} (\bibinfo{year}{2012}).

\bibitem[{\citenamefont{Kosowsky
  et~al.}(1992{\natexlab{a}})\citenamefont{Kosowsky, Turner, and
  Watkins}}]{Kosowsky:1992rz}
\bibinfo{author}{\bibfnamefont{A.}~\bibnamefont{Kosowsky}},
  \bibinfo{author}{\bibfnamefont{M.~S.} \bibnamefont{Turner}},
  \bibnamefont{and} \bibinfo{author}{\bibfnamefont{R.}~\bibnamefont{Watkins}},
  \bibinfo{journal}{Phys. Rev. Lett.} \textbf{\bibinfo{volume}{69}},
  \bibinfo{pages}{2026} (\bibinfo{year}{1992}{\natexlab{a}}).

\bibitem[{\citenamefont{Kamionkowski et~al.}(1994)\citenamefont{Kamionkowski,
  Kosowsky, and Turner}}]{Kamionkowski:1993fg}
\bibinfo{author}{\bibfnamefont{M.}~\bibnamefont{Kamionkowski}},
  \bibinfo{author}{\bibfnamefont{A.}~\bibnamefont{Kosowsky}}, \bibnamefont{and}
  \bibinfo{author}{\bibfnamefont{M.~S.} \bibnamefont{Turner}},
  \bibinfo{journal}{Phys. Rev.} \textbf{\bibinfo{volume}{D49}},
  \bibinfo{pages}{2837} (\bibinfo{year}{1994}), \eprint{astro-ph/9310044}.

\bibitem[{\citenamefont{Mandic and Buonanno}(2006)}]{PBBpaper}
\bibinfo{author}{\bibfnamefont{V.}~\bibnamefont{Mandic}} \bibnamefont{and}
  \bibinfo{author}{\bibfnamefont{A.}~\bibnamefont{Buonanno}},
  \bibinfo{journal}{Phys. Rev. D} \textbf{\bibinfo{volume}{73}},
  \bibinfo{pages}{063008} (\bibinfo{year}{2006}).

\bibitem[{\citenamefont{Guth}(1981)}]{guth}
\bibinfo{author}{\bibfnamefont{A.~H.} \bibnamefont{Guth}},
  \bibinfo{journal}{Phys. Rev. D} \textbf{\bibinfo{volume}{23}},
  \bibinfo{pages}{347} (\bibinfo{year}{1981}).

\bibitem[{\citenamefont{Ade et~al.}(2013)}]{Ade:2013zuv}
\bibinfo{author}{\bibfnamefont{P.}~\bibnamefont{Ade}} \bibnamefont{et~al.}
  (\bibinfo{collaboration}{Planck Collaboration}) (\bibinfo{year}{2013}),
  \eprint{1303.5076}.

\bibitem[{\citenamefont{Easther et~al.}(2007)\citenamefont{Easther, {Giblin
  Jr.}, and Lim}}]{Easther:2006vd}
\bibinfo{author}{\bibfnamefont{R.}~\bibnamefont{Easther}},
  \bibinfo{author}{\bibfnamefont{J.~T.} \bibnamefont{{Giblin Jr.}}},
  \bibnamefont{and} \bibinfo{author}{\bibfnamefont{E.~A.} \bibnamefont{Lim}},
  \bibinfo{journal}{Phys. Rev. Lett.} \textbf{\bibinfo{volume}{99}},
  \bibinfo{pages}{221301} (\bibinfo{year}{2007}), \eprint{astro-ph/0612294}.

\bibitem[{\citenamefont{Price and Siemens}(2008)}]{Price:2008hq}
\bibinfo{author}{\bibfnamefont{L.~R.} \bibnamefont{Price}} \bibnamefont{and}
  \bibinfo{author}{\bibfnamefont{X.}~\bibnamefont{Siemens}},
  \bibinfo{journal}{Phys. Rev. D} \textbf{\bibinfo{volume}{78}},
  \bibinfo{pages}{063541} (\bibinfo{year}{2008}).

\bibitem[{\citenamefont{Easther and Lim}(2006)}]{Easther:2006gt}
\bibinfo{author}{\bibfnamefont{R.}~\bibnamefont{Easther}} \bibnamefont{and}
  \bibinfo{author}{\bibfnamefont{E.~A.} \bibnamefont{Lim}},
  \bibinfo{journal}{JCAP} \textbf{\bibinfo{volume}{604}}, \bibinfo{pages}{010}
  (\bibinfo{year}{2006}).

\bibitem[{\citenamefont{Amin et~al.}(2014)\citenamefont{Amin, Hertzberg,
  Kaiser, and Karouby}}]{Amin:2014eta}
\bibinfo{author}{\bibfnamefont{M.~A.} \bibnamefont{Amin}},
  \bibinfo{author}{\bibfnamefont{M.~P.} \bibnamefont{Hertzberg}},
  \bibinfo{author}{\bibfnamefont{D.~I.} \bibnamefont{Kaiser}},
  \bibnamefont{and} \bibinfo{author}{\bibfnamefont{J.}~\bibnamefont{Karouby}}
  (\bibinfo{year}{2014}), \bibinfo{note}{\url{http://arxiv.org/abs/1410.3808}}.

\bibitem[{\citenamefont{Bin\'etruy et~al.}(2012)\citenamefont{Bin\'etruy,
  Boh\'e, Caprini, and Dufaux}}]{1201.0983}
\bibinfo{author}{\bibfnamefont{P.}~\bibnamefont{Bin\'etruy}},
  \bibinfo{author}{\bibfnamefont{A.}~\bibnamefont{Boh\'e}},
  \bibinfo{author}{\bibfnamefont{C.}~\bibnamefont{Caprini}}, \bibnamefont{and}
  \bibinfo{author}{\bibfnamefont{J.-F.} \bibnamefont{Dufaux}},
  \bibinfo{journal}{JCAP} \textbf{\bibinfo{volume}{06}}, \bibinfo{pages}{027}
  (\bibinfo{year}{2012}).

\bibitem[{\citenamefont{Dufaux}(2009)}]{0902.2574}
\bibinfo{author}{\bibfnamefont{J.~F.} \bibnamefont{Dufaux}},
  \bibinfo{journal}{Phys. Rev. Lett.} \textbf{\bibinfo{volume}{103}},
  \bibinfo{pages}{041301} (\bibinfo{year}{2009}).

\bibitem[{\citenamefont{Caprini et~al.}(2008)\citenamefont{Caprini, Durrer, and
  Servant}}]{0711.2593}
\bibinfo{author}{\bibfnamefont{C.}~\bibnamefont{Caprini}},
  \bibinfo{author}{\bibfnamefont{R.}~\bibnamefont{Durrer}}, \bibnamefont{and}
  \bibinfo{author}{\bibfnamefont{G.}~\bibnamefont{Servant}},
  \bibinfo{journal}{Phys. Rev. D} \textbf{\bibinfo{volume}{77}},
  \bibinfo{pages}{124015} (\bibinfo{year}{2008}).

\bibitem[{\citenamefont{Kosowsky
  et~al.}(1992{\natexlab{b}})\citenamefont{Kosowsky, Turner, and
  Watkins}}]{Kosowsky:1991ua}
\bibinfo{author}{\bibfnamefont{A.}~\bibnamefont{Kosowsky}},
  \bibinfo{author}{\bibfnamefont{M.~S.} \bibnamefont{Turner}},
  \bibnamefont{and} \bibinfo{author}{\bibfnamefont{R.}~\bibnamefont{Watkins}},
  \bibinfo{journal}{Phys. Rev.} \textbf{\bibinfo{volume}{D45}},
  \bibinfo{pages}{4514} (\bibinfo{year}{1992}{\natexlab{b}}).

\bibitem[{\citenamefont{Kosowsky and Turner}(1993)}]{Kosowsky:1992vn}
\bibinfo{author}{\bibfnamefont{A.}~\bibnamefont{Kosowsky}} \bibnamefont{and}
  \bibinfo{author}{\bibfnamefont{M.~S.} \bibnamefont{Turner}},
  \bibinfo{journal}{Phys. Rev.} \textbf{\bibinfo{volume}{D47}},
  \bibinfo{pages}{4372} (\bibinfo{year}{1993}), \eprint{astro-ph/9211004}.

\bibitem[{\citenamefont{Caprini and Durrer}(2006)}]{Caprini:2006jb}
\bibinfo{author}{\bibfnamefont{C.}~\bibnamefont{Caprini}} \bibnamefont{and}
  \bibinfo{author}{\bibfnamefont{R.}~\bibnamefont{Durrer}},
  \bibinfo{journal}{Phys. Rev.} \textbf{\bibinfo{volume}{D74}},
  \bibinfo{pages}{063521} (\bibinfo{year}{2006}), \eprint{astro-ph/0603476}.

\bibitem[{\citenamefont{Huber and Konstandin}(2008)}]{0709.2091}
\bibinfo{author}{\bibfnamefont{S.~J.} \bibnamefont{Huber}} \bibnamefont{and}
  \bibinfo{author}{\bibfnamefont{T.}~\bibnamefont{Konstandin}},
  \textbf{\bibinfo{volume}{05}}, \bibinfo{pages}{017} (\bibinfo{year}{2008}).

\bibitem[{\citenamefont{Abadie et~al.}(2010{\natexlab{b}})}]{s5_rates}
\bibinfo{author}{\bibfnamefont{J.}~\bibnamefont{Abadie}} \bibnamefont{et~al.}
  (\bibinfo{collaboration}{The LIGO Scientific and Virgo Collaborations}),
  \bibinfo{journal}{Classical Quantum Gravity} \textbf{\bibinfo{volume}{27}},
  \bibinfo{pages}{173001} (\bibinfo{year}{2010}{\natexlab{b}}).

\bibitem[{et(2013)}]{et}
\emph{\bibinfo{title}{Einstein telescope}} (\bibinfo{year}{2013}),
  \urlprefix\url{http://www.et-gw.eu/etsensitivities}.

\bibitem[{\citenamefont{Thrane and Romano}(2013)}]{locus}
\bibinfo{author}{\bibfnamefont{E.}~\bibnamefont{Thrane}} \bibnamefont{and}
  \bibinfo{author}{\bibfnamefont{J.~D.} \bibnamefont{Romano}},
  \bibinfo{journal}{Phys. Rev. D} \textbf{\bibinfo{volume}{88}},
  \bibinfo{pages}{124032} (\bibinfo{year}{2013}).

\bibitem[{\citenamefont{Phinney et~al.}(2004)}]{Phinney-et-al:2004}
\bibinfo{author}{\bibfnamefont{S.}~\bibnamefont{Phinney}} \bibnamefont{et~al.},
  \bibinfo{journal}{NASA Mission Concept Study}  (\bibinfo{year}{2004}).

\bibitem[{\citenamefont{Cutler and Harms}(2006)}]{Cutler-Harms:2006}
\bibinfo{author}{\bibfnamefont{C.}~\bibnamefont{Cutler}} \bibnamefont{and}
  \bibinfo{author}{\bibfnamefont{J.}~\bibnamefont{Harms}},
  \bibinfo{journal}{Phys. Rev. D} \textbf{\bibinfo{volume}{73}},
  \bibinfo{pages}{042001} (\bibinfo{year}{2006}).

\bibitem[{\citenamefont{Allen and Romano}(1999)}]{Allen-Romano:1999}
\bibinfo{author}{\bibfnamefont{B.}~\bibnamefont{Allen}} \bibnamefont{and}
  \bibinfo{author}{\bibfnamefont{J.~D.} \bibnamefont{Romano}},
  \bibinfo{journal}{Phys. Rev. D} \textbf{\bibinfo{volume}{59}},
  \bibinfo{pages}{102001} (\bibinfo{year}{1999}).

\bibitem[{\citenamefont{Thrane et~al.}(2013)\citenamefont{Thrane, Christensen,
  and Schofield}}]{schumann_ligo}
\bibinfo{author}{\bibfnamefont{E.}~\bibnamefont{Thrane}},
  \bibinfo{author}{\bibfnamefont{N.}~\bibnamefont{Christensen}},
  \bibnamefont{and} \bibinfo{author}{\bibfnamefont{R.~M.~S.}
  \bibnamefont{Schofield}}, \bibinfo{journal}{Phys. Rev. D}
  \textbf{\bibinfo{volume}{87}}, \bibinfo{pages}{123009}
  (\bibinfo{year}{2013}).

\bibitem[{\citenamefont{Mandic et~al.}(2012)\citenamefont{Mandic, Thrane,
  Giampanis, and Regimbau}}]{paramest}
\bibinfo{author}{\bibfnamefont{V.}~\bibnamefont{Mandic}},
  \bibinfo{author}{\bibfnamefont{E.}~\bibnamefont{Thrane}},
  \bibinfo{author}{\bibfnamefont{S.}~\bibnamefont{Giampanis}},
  \bibnamefont{and} \bibinfo{author}{\bibfnamefont{T.}~\bibnamefont{Regimbau}},
  \bibinfo{journal}{Phys. Rev. Lett.} \textbf{\bibinfo{volume}{109}},
  \bibinfo{pages}{171102} (\bibinfo{year}{2012}).

\bibitem[{\citenamefont{Lopez and Freese}(2013)}]{Lopez:2013mqa}
\bibinfo{author}{\bibfnamefont{A.}~\bibnamefont{Lopez}} \bibnamefont{and}
  \bibinfo{author}{\bibfnamefont{K.}~\bibnamefont{Freese}}
  (\bibinfo{year}{2013}), \eprint{1305.5855}.

\bibitem[{\citenamefont{Barnaby and Huang}(2009)}]{Barnaby:2009dd}
\bibinfo{author}{\bibfnamefont{N.}~\bibnamefont{Barnaby}} \bibnamefont{and}
  \bibinfo{author}{\bibfnamefont{Z.}~\bibnamefont{Huang}},
  \bibinfo{journal}{Phys. Rev. D} \textbf{\bibinfo{volume}{80}},
  \bibinfo{pages}{126018} (\bibinfo{year}{2009}), \eprint{0909.0751}.

\bibitem[{\citenamefont{Prokopec}(2001)}]{Prokopec:2001nc}
\bibinfo{author}{\bibfnamefont{T.}~\bibnamefont{Prokopec}}
  (\bibinfo{year}{2001}), \eprint{astro-ph/0106247}.

\bibitem[{\citenamefont{Anber and Sorbo}(2010)}]{Anber:2009ua}
\bibinfo{author}{\bibfnamefont{M.~M.} \bibnamefont{Anber}} \bibnamefont{and}
  \bibinfo{author}{\bibfnamefont{L.}~\bibnamefont{Sorbo}},
  \bibinfo{journal}{Phys. Rev.} \textbf{\bibinfo{volume}{D81}},
  \bibinfo{pages}{043534} (\bibinfo{year}{2010}), \eprint{0908.4089}.

\bibitem[{\citenamefont{Barnaby et~al.}(2011)\citenamefont{Barnaby, Namba, and
  Peloso}}]{Barnaby:2011vw}
\bibinfo{author}{\bibfnamefont{N.}~\bibnamefont{Barnaby}},
  \bibinfo{author}{\bibfnamefont{R.}~\bibnamefont{Namba}}, \bibnamefont{and}
  \bibinfo{author}{\bibfnamefont{M.}~\bibnamefont{Peloso}},
  \bibinfo{journal}{JCAP} \textbf{\bibinfo{volume}{1104}}, \bibinfo{pages}{009}
  (\bibinfo{year}{2011}), \eprint{1102.4333}.

\bibitem[{\citenamefont{Boyle and Buonanno}(2008)}]{Boyle:2007zx}
\bibinfo{author}{\bibfnamefont{L.~A.} \bibnamefont{Boyle}} \bibnamefont{and}
  \bibinfo{author}{\bibfnamefont{A.}~\bibnamefont{Buonanno}},
  \bibinfo{journal}{Phys. Rev. D} \textbf{\bibinfo{volume}{78}},
  \bibinfo{pages}{043531} (\bibinfo{year}{2008}), \eprint{0708.2279}.

\bibitem[{\citenamefont{de~Rham et~al.}(2012)\citenamefont{de~Rham, Gabadadze,
  and Tolley}}]{deRham:2011rn}
\bibinfo{author}{\bibfnamefont{C.}~\bibnamefont{de~Rham}},
  \bibinfo{author}{\bibfnamefont{G.}~\bibnamefont{Gabadadze}},
  \bibnamefont{and} \bibinfo{author}{\bibfnamefont{A.~J.}
  \bibnamefont{Tolley}}, \bibinfo{journal}{Phys. Lett.}
  \textbf{\bibinfo{volume}{B711}}, \bibinfo{pages}{190} (\bibinfo{year}{2012}),
  \eprint{1107.3820}.

\bibitem[{\citenamefont{de~Rham}(2014)}]{deRham:2014zqa}
\bibinfo{author}{\bibfnamefont{C.}~\bibnamefont{de~Rham}}
  (\bibinfo{year}{2014}), \eprint{1401.4173}.

\bibitem[{\citenamefont{Chamberlin and Siemens}(2012)}]{Chamberlin:2011ev}
\bibinfo{author}{\bibfnamefont{S.~J.} \bibnamefont{Chamberlin}}
  \bibnamefont{and} \bibinfo{author}{\bibfnamefont{X.}~\bibnamefont{Siemens}},
  \bibinfo{journal}{Phys. Rev.} \textbf{\bibinfo{volume}{D85}},
  \bibinfo{pages}{082001} (\bibinfo{year}{2012}), \eprint{1111.5661}.

\bibitem[{\citenamefont{Kibble}(1976)}]{Kibble:1976sj}
\bibinfo{author}{\bibfnamefont{T.}~\bibnamefont{Kibble}},
  \bibinfo{journal}{J.Phys.} \textbf{\bibinfo{volume}{A9}},
  \bibinfo{pages}{1387} (\bibinfo{year}{1976}).

\bibitem[{\citenamefont{Sendra and Smith}(2012)}]{sendra}
\bibinfo{author}{\bibfnamefont{I.}~\bibnamefont{Sendra}} \bibnamefont{and}
  \bibinfo{author}{\bibfnamefont{T.~L.} \bibnamefont{Smith}},
  \bibinfo{journal}{Phys. Rev. D} \textbf{\bibinfo{volume}{85}},
  \bibinfo{pages}{123002} (\bibinfo{year}{2012}).

\bibitem[{\citenamefont{Allen}(1997)}]{allen97}
\bibinfo{author}{\bibfnamefont{B.}~\bibnamefont{Allen}}, in
  \emph{\bibinfo{booktitle}{Les Houches School of Physics: Astrophysical
  Sources of Gravitational Radiation}}, edited by
  \bibinfo{editor}{\bibfnamefont{J.-P.~L.} \bibnamefont{J.-A.~Marck}}
  (\bibinfo{publisher}{Cambridge Contemporary Astrophysics},
  \bibinfo{year}{1997}), p. \bibinfo{pages}{373}.

\end{thebibliography}

\end{document}